

\magnification\magstep1
\hsize 27pc
\vsize 43pc
\nopagenumbers
\headline{\ifodd\pageno\rightheadline \else\leftheadline\fi}
\footline{\hfil}

\voffset .2 truein
\hoffset .55truein

\font\fb=cmr12
\font\fc=cmr10
\font\fab=cmr9
\font\fs=cmr6
\font\fd=cmr5
\font\bfab= cmbx9

\def\eightpoint{\def\rm{\fam0\eightrm}
  \font\eightrm=cmr9
  \font\sixrm=cmr6
  \font\fiverm=cmr5
  \font\eighti=cmmi9
  \font\sixi=cmmi6
  \font\fivei=cmmi5
  \font\eightsy=cmsy9
  \font\sixsy=cmsy6
  \font\fivesy=cmsy5
  \font\eightbf=cmbx9
		\font\sixbf=cmbx6
  \font\fivebf=cmbx5
  \font\eightit=cmti9
  \font\eighttt=cmtt9
  \font\eightsl=cmsl9
  \font\tentex=cmtex10
  \font\sevenrm=cmr7
  \textfont0 = \eightrm
  \scriptfont0 = \sixrm
  \scriptscriptfont0 = \fiverm
  \textfont1 = \eighti
  \scriptfont1 = \sixi
  \scriptscriptfont1 = \fivei
  \textfont2 = \eightsy
  \scriptfont2 = \sixsy
  \scriptscriptfont2 = \fivesy
  \textfont3=\tenex
  \scriptfont3=\tenex
  \scriptscriptfont3=\tenex
  \textfont\itfam=\eightit  \def\it{\fam\itfam\eightit}%
  \textfont\slfam=\eightsl  \def\sl{\fam\slfam\eightsl}%
  \textfont\ttfam=\eighttt  \def\tt{\fam\ttfam\eighttt}%
  \textfont\bffam=\eightbf  \scriptfont\bffam=\sixbf
  \scriptscriptfont\bffam=\fivebf  \def\bf{\fam\bffam\eightbf}%
  \normalbaselineskip=11pt
  \setbox\strutbox=\hbox{\vrule height8pt depth3pt width0pt}%
  \let\sc=\sevenrm  \let\big=\eightbig \normalbaselines\rm}

 \def\eightbig#1{{\hbox{$\textfont0=\tenrm\textfont2=\tensy
   \left#1\vbox to7.25pt{}\right.\n@space$}}}

\def\ninepoint{\def\rm{\fam0\ninerm}
  \font\ninerm=cmr9
  \font\sixrm=cmr6
  \font\fiverm=cmr5
  \font\ninei=cmmi9
  \font\sixi=cmmi6
  \font\fivei=cmmi5
  \font\ninesy=cmsy9
  \font\sixsy=cmsy6
  \font\fivesy=cmsy5
  \font\ninebf=cmbx9
		\font\sixbf=cmbx6
  \font\fivebf=cmbx5
  \font\nineit=cmti9
  \font\ninett=cmtt9
  \font\ninesl=cmsl9
  \font\tentex=cmtex10
  \font\sevenrm=cmr7
  \textfont0 = \ninerm
  \scriptfont0 = \sixrm
  \scriptscriptfont0 = \fiverm
  \textfont1 = \ninei
  \scriptfont1 = \sixi
  \scriptscriptfont1 = \fivei
  \textfont2 = \ninesy
  \scriptfont2 = \sixsy
  \scriptscriptfont2 = \fivesy
  \textfont3=\tenex
  \scriptfont3=\tenex
  \scriptscriptfont3=\tenex
  \textfont\itfam=\nineit  \def\it{\fam\itfam\nineit}%
  \textfont\slfam=\ninesl  \def\sl{\fam\slfam\ninesl}%
  \textfont\ttfam=\ninett  \def\tt{\fam\ttfam\ninett}%
  \textfont\bffam=\ninebf  \scriptfont\bffam=\sixbf
  \scriptscriptfont\bffam=\fivebf  \def\bf{\fam\bffam\ninebf}%
  \normalbaselineskip=11pt
  \setbox\strutbox=\hbox{\vrule height8pt depth3pt width0pt}%
  \let\sc=\sevenrm  \let\big=\ninebig \normalbaselines\rm}

 \def\ninebig#1{{\hbox{$\textfont0=\tenrm\textfont2=\tensy
   \left#1\vbox to7.25pt{}\right.\n@space$}}}


\def\abs#1{\bigskip\bigskip\medskip{\ninepoint%
       {\narrower \noindent%
      {\bf Abstract.} #1 \medskip\bigskip}}\medskip}

\def\Acknowledgments#1{\goodbreak\bigskip\noindent{%
{\bf Acknowledgments.} #1.}}  


\def\ats{\kern.15em\hbox{@}\kern.15em}

\def\aut{\bigskip \bigskip
        \centerline {\author} }

\def\CC{\hbox{\rm C\kern -.58em {\raise .54ex \hbox{$\scriptscriptstyle |$}}
  \kern-.55em {\raise .53ex \hbox{$\scriptscriptstyle |$}} }}

\def\copyright{\hbox{{\fb o}\kern-.61em \raise .46ex \hbox{\fd c}}}

\def\cor#1 #2{\begingroup\medbreak\noindent{\bfab Corollary #1 }\sl #2}

\def\defin#1 #2{\begingroup\medbreak\noindent{\bfab Definition #1 } #2}

\rm

\rm



\def\footnoterule{\kern -3pt \hrule width 0truein \kern 2.6pt}

\def\frac#1#2{{#1 \over #2}}

\def\lem#1 #2{\begingroup\medbreak\noindent{\bfab Lemma #1 }\sl #2}

\def\mbf#1{\setbox0=\hbox{#1}%
           \kern-.025em\copy0\kern-\wd0
           \kern.05em\copy0\kern-\wd0
           \kern-.025em\hbox{\raise.0433em\box0} }

\def\NN{{{\rm l}\kern-.15em{\rm N}}}

\def\nullx{\hfill}

\def\per{\kern .03em \% \kern .02em}





\def\pro#1 #2{\begingroup\medbreak\noindent{\bfab Proposition #1 }\sl #2}

\def\References{\goodbreak\bigskip\centerline{\bf References}\medskip
   \frenchspacing \hfuzz 2pt}

\rm

\def\rightheadline{\ifnum\pageno=\count100 \nullx%
  \else\it\chptitle\hfil\rm\folio\fi}
  \def\leftheadline{\ifnum\pageno=\count100 \nullx%
  \else\rm\folio\hfil\it\authead\fi}  

\def\RR{{{\rm l}\kern-.17em{\rm R}}}

\def\sRR{{\sl \hbox{I\kern-.2em\hbox{R}}}}

\def\sect#1{\goodbreak\bigskip\smallskip\centerline{\bf\S
#1}\bigskip\noindent\ignorespaces}


\def\tabone#1 #2{\goodbreak\medskip\centerline{\bfab Table #1.}\smallskip
  \centerline {#2.} \smallskip}

\def\tabtwo#1 #2 #3{\goodbreak\medskip\centerline{\bfab Table #1.}\smallskip
   \centerline {#2} \vskip.1pt
    \centerline {#3.}\smallskip}

\def\tabthre#1 #2 #3 #4{\goodbreak\medskip\centerline{\bfab Table
   #1.}\smallskip
   \centerline {#2} \vskip.1pt
   \centerline {#3} \vskip.1pt
   \centerline {#4.}\smallskip}

\def\theo#1 #2{\begingroup\medbreak\noindent{\bfab Theorem #1 }\sl #2}
 \def\tit#1{\centerline {\bf #1}}
\def\titwo#1{\medskip \centerline {\bf #1}}

\def\titexp#1#2{\hbox{{\bf #1} \kern-.25em \raise .90ex \hbox{\bf #2}}\/}
\def\titsub#1#2{\hbox{{\bf #1} \kern-.25em \lower .60ex \hbox{\bf #2}}\/}

\def\ZZ{{{\rm Z}\kern-.52em{\rm Z}}}




\centerline{}
\pageno=\count100
\count102=\count100
\advance\count102 by \count101
\advance\count102 by -1
\insert\footins{\fs
\medskip
\baselineskip 8pt
\leftline{{Approximation Theory VIII}
\hfill {\fc \the\pageno}}
\leftline{Charles K. Chui and Larry L. Schumaker\ (eds.),
    pp. \the\pageno--\the\count102.}
\leftline{Copyright \copyright\ 1995 by World Scientific Publishing Co., Inc.}
\leftline{All rights of reproduction in any form reserved.}
\leftline{ISBN 0-12-xxxxxx-x}
\smallskip
\par\allowbreak}

\def\ref{\global\advance\refnum by 1 \item{\the\refnum .}}
\newcount\refnum \refnum = 0

\def\em{\it}

\def\Qbf{{\bf Q}\,}

\def\hh{\hat{h}}
\def\fb{\overline{f}}
\def\ft{\tilde{f}}
\def\zt{\tilde{z}}
\def\kap{{\kappa}}
\def\th{{\theta}}
\def\thh{\hat{\theta}}
\def\part{\partial}
\def\Sh{\hat{S}}
\def\Shb{\bar{\hat{S}}}
\def\fqSh{\widehat{\part_{f}^qS}}
\def\fqthh{\widehat{\part_{f}^q\th}}

\def\nubf{{ \nu}}
\tit{Adaptive Kernel Estimation of the Spectral} 
\titwo{Density with Boundary Kernel Analysis}
\def\chptitle{Boundary Kernel Spectral Estimation}
\def\author{Alexander Sidorenko and Kurt S.~Riedel}
\def\authead{A.~Sidorenko and K.S.~Riedel}
{\vskip-.12in}
\aut
{\vskip-.285in}
\abs{A hybrid estimator of the log-spectral density of a stationary time series
is proposed. First, a multiple taper estimate is performed, followed by
kernel smoothing the log-multitaper estimate. This procedure reduces the
expected mean square error by $({\pi^2 \over 4})^{.8}$ over simply
smoothing the log tapered periodogram. The optimal number of tapers is $O(N^{
8/15})$. A data adaptive implementation of a variable bandwidth kernel
smoother is given. When the spectral density is discontinuous,
one sided smoothing estimates are used.}
{\vskip-.425in}
\sect{1. 
Introduction}
We consider a discrete, stationary, Gaussian, time
series $\{ x_j , j=1, \ldots N \}$
with a piecewise smooth spectral density, $S(f)$, that is bounded away from
zero.
The autocovariance is the Fourier transform of the spectral density:
Cov $[x_j ,x_k ] = \int_{-1/2}^{1/2} S(f)e^{2\pi i(j-k)f} df$.
Our goal is to estimate the log spectral density:
$\theta (f) \equiv \ln [S(f)]$.
We consider data adaptive kernel smoother estimators which
self-consistently estimate the best local halfwidth for smoothing. 
When the spectral density is discontinuous at a point,
the kernel estimate must be one-sided. We present new ``boundary''
kernels for one-sided estimation. 
{\vskip-.1in}
\sect{2. 
Variance of 
multitapering and kernel smoothing}
{\vskip-.1in}
The multiple taper estimate of the spectral density is
a quadratic estimate of the form:
{\vskip-.18in}
$$
\hat{S}_{MT} (f) =  \sum_{m,n=1}^N Q_{mn} x_m x_n e^{2 i (m-n)f} = 
\sum_{k=1}^K \mu_k \left| \sum_{n=1}^N \nu_n^{(k)}
x_n e^{-2 \pi inf} \right|^2 \  
 , \eqno (1)
$$
where the $\mu_k$ and $\nubf^{(k)}$ are 
$K$ eigenvalues and orthonormal eigenvectors  of $\Qbf$. 
In the large $N$ limit,
the sinusoidal tapers are optimal: $\nu_m^{(k)} = \sqrt{{2 \over N+1}}
\sin \left({\pi k m \over N+1}\right) $ [4]. Let $\mu_k \equiv 1/K$.
For these tapers, the spectral estimate (1) reduces to  
{\vskip-.12in}
$$
\hat{S}_{MT} (f) = {\Delta\over K} \sum_{k=1}^K 
 |y(f+k \Delta )-y(f-k \Delta )|^2 \ ,\eqno (2)$$
where $y(f)$ is the FT of $\{ x \}$: $y(f) = \sum_{m=1}^N x_m e^{-2 \pi imf}$
and $\Delta \equiv 1/(2N+2)$.
The sinusoidal multitaper estimate reduces the bias since the
sidelobes of $y(f+k \Delta )$ are partially cancelled by those of
$y(f-k \Delta )$.

 To leading order in $K/N$, the local white noise approximation \hbox{holds:}
$\Sh_{MT}(f)$ has a $\chi^2_{2K}$ distribution. 
 Note $
{\bf E  }\left[   \ln \left( \chi^2_{2K} \right) \right] =
\psi (K)  - \ln(K)$,
${\bf Var }\left[   \ln \left( \chi^2_{2K} \right) \right] =\   \psi'(K)
$,
where $\psi$ is the digamma function. 

We  now consider kernel estimators of the $q$th derivative,
$\partial_f^{q}S(f)$:
$$
\widehat{\partial_f^{q}S_{}} (f) \equiv \ {1\over h^{q+1}}
\int_{-1/2}^{1/2} \kappa \left( {f^{\prime} -f
\over h} \right) \hat{S}_{MT} (f^{\prime} ) df' \ ,\eqno (3)
$$
where
$\kappa (f)$ is a smooth kernel with support in $[-1,1]$, and
$\kappa ( \pm 1)=0$ and $h$ is the bandwidth parameter.
We say a kernel
is of order $(q,p)$ if
$\int f^m \kap(f)df =  { q!\ }\delta_{m,q} \ , \ m= 0, \ldots,p-1$.
We denote the $p$th moment of a kernel of order $(q,p)$ by $B_{q,p}$:
$\int f^p \kap(f)df  =   p!\ B_{q,p}$.
For function estimation ($q=0$), we  use $p=2$ and $p=4$. 
To estimate the second derivative,
we use a kernel of order (2,4).

The estimator (3) is a quadratic estimator as in (1) with
$Q_{mn} = \hat{\kappa}_{m-n} \sum_{k=1}^K  \nu_m^{(k)}\nu_n^{(k)}$,
where $\hat{\kappa}_m \equiv h^{-(q+1)}\int \kappa ({f' \over h})
e^{imf'} df'$ is the FT of the kernel.  
We evaluate the variance of (3),
using the local white noise approximation: 
{\vskip-.22in}
$$
{\bf Var} \left[ \widehat{\partial_f^qS}_{} (f)\right]/ S(f)^2 = 
{\rm tr} [{\bf QQ}] 
= 
\sum_{k,k'=1}^K  \sum_{n=1}^N \sum_{m=1-n}^{N-n} \tilde{\kappa}_m^2
\nu_{n+m}^{(k)} \nu_n^{(k)} \nu_{n+m}^{(k')} \nu_n^{(k')} \
 \eqno (4)$$
{\vskip.05in}
For $mh \gg 1$, note that $\hat{\kappa}_m \sim\ 
{\cal O}(\|\hat{\kap} \| /(mh)^2 )$.
Since the $k$-th taper has a scale length of variation of $N/k$,
$\nu_{n+m}^{(k)} {\simeq}\ \nu_n^{(k)} [1+\ {\cal O}({km \over N})]$.
We expand (4) in $mk/N$ for $|mh |<{\cal O}((Nh/K)^{1/5})$ and
drop all terms with $|mh|>{\cal O}((Nh/K)^{1/4})$:
{\vskip-.12in}
$${\bf Var} \left[\fqSh(f)\right]\ \underline{\sim}\ 
S(f)^2 \sum_{k,k'=1}^K  \left( \sum_{n=1}^N
| \nu_n^{(k)}|^2 | \nu_n^{(k')} |^2 \right) \left(
\sum_{m=1}^N \tilde{\kappa}_m^2 \right) \ . \eqno (5)
$$
{\vskip-.1in}
The last line is valid to ${\cal O}(1/(mh)^4 ) + {\cal O}({Km / N})$,
yielding a final accuracy of (5) of ${\cal O}((K/Nh)^{4/5})$. 
For the sinusoidal tapers, (5) reduces to
{\vskip-.08in}
$${\bf Var} \left[ \fqSh(f)\right] \underline{\sim}
{\|{\kappa}^2\|S(f)^2 \over Nh^{(q+1)} }\left|1+ {1\over 2K}\right|
+\ {\cal O}(({K \over Nh})^{4/5}) \ ,
$$
{\vskip-.06in} \noindent
where $\|{\kappa}^2\|$ is the square integral of $\kappa$.
To calculate the local bias, we
Taylor expand the spectral density. 
Note $\int_{-1/2}^{1/2} |f^{\prime} |^2| V^{(k)} (f')|^2 df' =
k^2/(4N^2)$.        
where $V^{(k)}$ is the FT  of   
the sinusoidal $\nubf^{(k)}$. 
{\vskip -.08in}
\sect
{3. Smoothed log-multitaper estimate}
{\vskip -.08in}
We define the multitaper estimate of the logarithm of the spectral density by
$\thh_{MT}(f) \equiv\ {\rm ln}[\Sh_{MT}(f)] - [\psi (K) -  \ln(K)]/K$. 
By averaging the $K$ estimates prior to taking the
logarithm, we reduce both the bias and the variance. 
The variance reduction factor from using ${\ln[\Shb(f)]}$,
instead of $\overline{\ln[\Sh(f)]}$ is $K\psi'(K)/\psi'(1)$. 
For large $K$, $K\psi'(K) \simeq 1 + \frac{1}{2K}$, so the variance
reduction factor is asymptotically $6/\pi^2$.   

We define the smoothed multitaper ln-spectral estimate:
$\widehat{\partial_f^q\theta}_{\kappa} (f)$ by kernel smoothing
$\thh_{MT}(f)$ analogous to Eq.~(3).
To evaluate the error in
 $\widehat{\partial_f^q\theta}_{\kappa} (f)$, 
we expand
$\hat{\theta}_{MT}(f')$ about $\theta (f)$: $\hat{\theta}_{MT}
(f^{\prime} ) \ {\simeq}\ \ln [ \theta (f)]$
$+ [ \hat{S}_{MT}(f) -S(f^{\prime} )] / S(f^{\prime})$,
that is valid when $K \ll N$,
$h \ll 1$, and $Nh \gg 1$. 
The leading order bias is
{\vskip-.18in}
$$
{\bf Bias} [ \widehat{\partial_f^q\theta} (f)] 
\simeq B_{q,p} \partial_f^p \theta(f) h^{p-q }  +
\partial_f^q[ \theta'' + | \theta^{\prime} |^2 (f)] {K^2 \over 24N^2}
\ .\eqno(6)
$$
The first term is the bias from kernel smoothing
and the second term is from the sinusoidal multitaper estimate.
To leading order in $1/K$,
the variance of $\fqthh(f)$ reduces to
the same calculation as the variance of $\fqSh(f)$.
To this order, the variance inflation factor from
the long tail of the $\ln[\chi^2_{2K}]$ distribution is not visible.
Since ${\bf Cov} [ \hat{\th}_{MT} (f'),\hat{\th}_{MT} (f'')]
\le [\psi'(K)/K] \times {\bf Cov} [ \hat{S}_{MT} (f'),\hat{S}_{MT} (f'')]$,
we have 
{\vskip -.06in}
$$
{\bf Var} \left[ \widehat{\partial_f^q\th}(f)\right]  \approx
{\|{\kappa}\|^2 \over Nh^{2q+1} }\left|1+ {1\over 2K}\right|^2
\ .\eqno (7)
$$
{\vskip -.12in} \noindent
Using (6) and (7), the expected asymptotic square error (EASE) in 
$\widehat{\partial_f^q \th}$:
{\vskip-.08in}
$$ {\bf E}\left[\left|\widehat{\partial_f^q \th}(f_j) -
{\partial_f^q \th}(f_j) \right|^2\right] \ \approx \
{\bf Var} \left[ \widehat{\partial_f^q\th}(f)\right] \ + \
{\bf Bias}^2 [ \widehat{\partial_f^q\theta} (f)] 
\ ,\eqno (8)
$$ 
where
corrections are ${\cal O}(h^{2(p-q)+1}) 
+ {\cal O}({1 \over K})+\ {\cal O}(({K \over Nh})^{4/5})$ 
$+ {\cal O}({1\over Nh})$.
The benefit of multitapering (in terms of the  
variance reduction) tends rapidly to zero. 
Minimizing (8) with respect to $h$ and $K$ yields
$K_{opt} << Nh_{opt}$ and that to leading order
{\vskip-.12in}
$$h_{o}(f) = \left[ {2q+1 \over 2(p-q)}
{  \|\kap\|^2 \over  B_{q,p}^2 N |\partial_f^p \th(f_j)|^2 }
\left|1+{1\over 2K}\right|^2
\right]^{1\over 2p+1}
\ ,\eqno(9)$$
and  $K_{opt} \sim N^{(3p+q +2)/(6p+3)}$.
For kernels of order $(0,2)$, this reduces to
$h_{opt} \sim N^{-1/5}$ and  $K_{opt} \sim N^{8/15}$.
Thus the ordering $1 \ll K \ll Nh$ is justified.
The EASE depends only weakly on $K$ for
$1 \ll K \ll Nh$ while the dependence on the choice of bandwidth is strong.
When the bandwidth, $h_o$, satisfies (9),
the leading order EASE  reduces to
{\vskip-.08in}
$$
{\bf E}\left[\left|\widehat{\partial_f^q \th}(f_j) -
{\partial_f^q \th}(f_j) \right|^2\right]  \sim
  |B_{q,p} \partial_f^p \th(f_j)|^{(4q+2)\over(2p+1)}
\left({  \|\kap\|^2  \over N } \right)^{2(p-q)\over(2p+1)}.
 \eqno(10)$$
Thus the EASE in estimating
$\partial_f^q \th$ is proportional to $N^{-2(p-q)\over(2p+1)}$.
The loss depends on the kernel shape through  $B_{qp}(\kappa)$ and 
$ \|\kap\|^2$. In Sec.~5, we optimize the kernel shape subject to
moment constraints.
If $K=1$ (a single
taper), the variance term in (7) should be inflated by a factor of
${\pi^2 \over 6} \sum_{n=1}^N \nu_n^4$. Thus using a moderate level of
multitapering prior to smoothing the logarithm reduces the EASE by a
factor of $[{\pi^2 \over 6} \sum_{n=1}^N \nu_n^4 ]^{4/5}\
=\ [{\pi^2 \over 4}]^{4/5}$, where we use
$ \sum_{n=1}^N \nu_n^4 =\ 1.5 $. 
{\vskip -.12in}
\sect{4. 
Data adaptive estimate}
In practice, $\theta^{\prime\prime}(f)$ is unknown and we use a data adaptive
multiple stage kernel estimator where a pilot estimate of the optimal
bandwidth is performed
prior to estimating $\theta (f)$. When the spectral
range is large, it is often essential to allow the bandwidth 
to vary locally as a function of frequency.
If computational effort is not important,
set $K=N^{8/15}$; otherwise we
set $K$ by the computational budget.
For nonparametric function estimation, data adaptive multiple stage schemes 
are given in [1-5].
Our scheme for spectral estimation has the following steps:

\noindent
0) Compute the Fourier transform, $y(f)$ on a grid of size $2N+2$ and
$\hat{\theta}_1 (f)=\ln [|y(f+ \Delta )-y(f-\Delta )|^2 /2(N+1)]$ on a
grid of size $N+1$.
Evaluate $\thh_{MT}(f)$ on a grid of size $2N+2$.

\noindent
1) Smooth the tapered log-periodogram,  $\hat{\theta}_1 (f)$,
a kernel of order (0,4). Choose the global halfwidth by either
the Rice criterion [1] or by fitting the average square residual
as a function of the global halfwidth to a two parameter model [3].

\noindent
2) Estimate $\theta''(f)$ by kernel smoothing the multitaper estimate
$\thh_{MT}$ with global halfwidth $h_{2,4}$.
Relate the optimal (2,4) to  the optimal (0,4) global halfwidth
using the halfwidth quotient relation [1,5]:
$h_{2,4} =H(\kap_{2,4},\kap_{0,4}) h_{0,4}$.

\noindent
3) Estimate $\theta(f)$ using the variable halfwidth given
by substituting $\hat{\theta}''(f)$
into the optimal halfwidth expression of Eq.~(9).

An estimation scheme has a relative 
convergence rate of $N^{-\alpha}$ if
{\vskip-.08in}
$${\bf E}\left[|\hat{\th}(f|\hh_{0,2}) - \th(f)|^2 \right] \ \simeq \  
\left( 1+ {\cal O}(C_r^2 N^{-2\alpha})\right)
{\bf E}\left[|\hat{\th}(f|h_{0,2}) - \th(f)|^2\right]
\ ,
$$
{\vskip-.08in}
\noindent
where $h_{0,2}$ is the optimal halfwidth and $\hat{h}_{0,2}$ is the
estimated halfwidth.
Our data-adaptive method
has a convergence rate of $N^{-4/5}$ and
a relative  convergence rate: $N^{-2/9}$.

If multitapering were used in step 1, then $\theta(f_n )$
would be correlated at neighboring Fourier frequency. The multitaper
induced autocorrelation would then bias the estimate of the $h_{o,4}$.
halfwidth. 
We use a single taper in Step 1 and many tapers in Steps 2-3.
To correct for this, we 
inflate the variance in the (0,4) kernel estimate. 
The halfwidth quotient relation
relates the optimal halfwidth for derivative estimates,
$\hh_{2,4}$ to that of the 
$(0,4)$ kernel using (9):
$ \hh_{2,4} = H(\kap_{2,4},\kap_{0,4}) \hat{h}_{0,4}$ , \ \ 
{\rm where} \  
{\vskip-.16in}
$$
H(\kap_{2,4},\kap_{0,4}) \equiv 
\left({ 10 B_{0,4}^2 \|\kap_{2,4}\|^2\over
B_{2,4}^2 \|\kap_{0,4}\|^2 }\right)^{1\over 9}
\left({ \pi^2 \sum_n |\nu^{(1)}_n|^4\over 6 } \right)^{1\over 9}.
$$
{\vskip-.16in}
When $\hat{\theta}^{\prime\prime}(f)$ is vanishingly small, the optimal
halfwidth becomes large. Thus, $\hh_{0,2}$ needs to be regularized. 
Following Riedel \& Sidorenko (1994), we determine the size of the
regularization from $\hat{h}_{0,4}$ in the previous stage.
{\vskip-.08in}
\sect{5.
Discontinuities and boundary kernels}
{\vskip-.06in}
When $S(f)$ or its derivatives are discontinuous, the kernel estimate
needs to be one sided using only the data to the left (or right) of
discontinuity, $f_{disc}$. A similar problem occurs when we wish 
to estimate the spectral density near the boundary, in our case at
$f=1/2$ or $f=0$ if $S'(0) \ne 0$. In these cases, the estimation
point, $f$ is not in the center of the kernel domain, that we
denote as $f_i \in [f_{disc}, f_{disc} +2h]$.
The kernel estimate is a weighted average of the $\thh_i$: 
$\widehat{\partial_f^q \th}(f) = \sum_i K(f,f_i)\thh_i $
in the frequency interval, 
$f_i \in [f_{disc}, f_{disc} +2h]$.
The kernel function, $K(f,f_i)$ must still satisfy the moment conditions:
$\sum_i K(f,f_i)(f_i-f)^j= q!  \delta_{j,q}$, $0\le j <m$.
Thus $K(f,f_i)$ is asymmetric and cannot be simply a function of $f-f_i$.
We now describe results in [6] on kernel estimation near a boundary or
discontinuity.

We assume that the estimation point, $f$ satisfies $f \ge f_{disc}$.
Far  from the discontinuity, we use the standard kernel estimate (3)
with $h_o(f)$ given by (9). 
As $f$ approaches $f_{disc}$, the domain of the kernel touches
the discontinuity when $f -f_{disc} = h_o(f)$.
This point is the start of the boundary region around the discontinuity.
We call this point,  ``the right touch point'', $f_{tp}$ as defined by
the equation: $f_{tp} -f_{disc} = h_o(f_{tp})$. 
In the boundary region between
$f_{disc}$ and $f_{tp}$, we use a fixed halfwidth, $h$ and modify
the kernel shape to satisfy the moment conditions.

We  define $\fb\equiv  f_{disc} +\ h$, $\ft\equiv (f- f_{disc})/ h$.
The ``measurements'' are the log multitaper values, that are evaluated
on a grid of size $2N$: $\thh_i \equiv \thh_{MT}(f_i\equiv i/2N)$
We standardize the grid points: $\zt_i  \equiv (f_i- f_{disc})/ h$. 
The kernel estimate is a weighted average of the $\thh_i$ 
in the frequency interval, 
$f_i \in [f_{disc}, f_{disc} +2h]$. The total halfwidth is $2h$
. We define orthonormal
polynomials, $P_j$ on $[f_{disc}, f_{disc} +2h]$ by
$\frac{1}{Nh} \sum_{i}P_k(\zt_i)P_j(\zt_i)   = g_k \delta_{kj}$
where $g_k$ is a normalization.
We have expanded the kernel function, $ K(f,f_i)$ in
the polynomials, $P_j$:
{\vskip-.08in}
$$
\widehat{\partial_f^q \th}(f) = \sum_i K(f,f_i)\thh_i 
= {1 \over Nh^{q+1}}
\sum_i \sum_j b_j(f) P_j(\zt_i)\thh_i
.$$
{\vskip-.1in}
The moment conditions become
$\sum_k C_{kj} b_k = \delta_{qj}$ for $j=0,\ldots ,p-1$,
where
$C_{kj}(f) \equiv \sum_{i} 
  P_k(\zt_i)        
 \frac{1}{j!} (-\ft)^j$.
The matrix $C_{kj}(f)$ is upper triangular.
We solve for $b_0, b_1 \ldots  b_{p-1}$:
{\vskip -.24in}
$$ 
 b_j  =  \frac{\delta_{j,q}}{C_{qq}}  \;\;\;
 {\rm for } \;\; 0\le j\le q \; ;\ 
{\rm and} \ \
b_j  =  - \frac{1}{C_{jj}}\sum_{i=q}^{j-1} C_{ij}b_i
                  \;\; {\rm for\ } \; q< j <p 
\ . $$
{\vskip -.06in}
Thus the  moment conditions  prescribe 
$b_0, b_1 \ldots  b_{p-1}$
while the coefficients $b_p,b_{p+1},\ldots$ are free parameters.
The leading order bias equals
$\th^{(p)}(f) h^{p-q}$ $ \sum_{k=q}^p C_{kp} b_k$.
The summation stops at $k=p$ because $C_{kp}=0$ for $k>p$.
The EASE is a quadratic function of $b_j$:
{\vskip .1in}

\centerline{$ 
 EASE(f)  =  \frac{1}{N h^{2q+1}}
 \sum_{k\geq q} g_k b_k^2   +
 \left( \th^{(p)}(f) h^{p-q} \sum_{k=q}^p C_{kp} b_k\right)^2
 \; .
$} 
{\vskip .1in}

\noindent
In the absence of boundary conditions,
EASE attains the minimum when $b_k=0$ for $k>p$.
and the optimal value $b_p$ can be easily found.

We now restrict to the continuum limit when $Nh \rightarrow \infty$
and the points are equispaced. In this case, the discrete sums
become integrals and the $P_j$ become Legendre functions.
We also require $p = q+2$.
We seek a boundary kernel in the form:
$
 K(f,f_i)=\frac{\gamma_q}{ h ^{q+1}}
        G\left(\ft,\zt_i \right)$,
where $ \gamma_q = \frac{1}{2} \prod_{k=1}^{q}(2k+1) $.
The function $G(\ft,\zt)$ is
the {\em normalized boundary kernel},
and its domain is  $\ft\in [-1,0]$, $\zt\in [-1,1]$.
%
Using the Legendre polynomials, $P_j$,
we expand the normalized boundary kernel:
$ G(\ft,\zt) =  \sum_j b_j(\ft) P_j(\zt).$

We parameterize the kernel shape by $\beta =\frac{h}{h_0(f)}$ 
where $h$ is the actual halfwidth and ${h_0(f)}$ is given in Eq.~(9). 
Using $\beta$ instead of
$\th^{(p)}(f)$ is advantageous 
because we are interested in kernels that have a fixed
halfwidth, $h$ in the boundary region: 
$h(f)=h_0(f_{tp})$ and $\beta =\frac{h_0(f_{tp})}{h_0(f)}$

\noindent {\bf Theorem.}
{\em
Among all boundary kernels with support $[0,2h]$,
the kernel which minimizes the  leading order EASE is
$K(f,f_i)=\frac{\gamma_q}{h^{q+1}}G\left(\ft,\zt_i\right)$
where
$$ 
G(\ft,\zt_i) = 
 P_q(\zt_i)
\; + \; (2q+3) \ft
 P_{q+1}(\zt_i)
\; + \;
 \frac{(2q+3)\ft^2-1}
{\frac{2q+3}{(2q+5)\beta^{2q+5}} + \frac{2}{2q+5}}
 P_{q+2}(\zt_i)
\; ,
$$
$P_q, P_{q+1}, P_{q+2}$ are the Legendre polynomials,
$\beta=h/h_0(f)$ [6].
}

For $h(f)=h_0(f)$, the optimal boundary kernel
simplifies to
$$
G(\ft,\zt) \; = \;
 P_q(\zt)
\; + \; (2q+3) \ft
 P_{q+1}(\zt)
 \; + \;
         ( (2q+3)\ft^2 -1 )
 P_{q+2}(\zt)
\; .$$
At the touch point, $f= f_{tp} \equiv f_{disc} + h(f_{tp})$ with
$h=h_0(f_{tp})$, the optimal boundary kernel is identical to
the optimal interior kernel:
{\vskip -.1in}
$$G(\ft,\zt) \; = \;
 P_q(\ft-\zt)-P_{q+2}(\ft-\zt)
\ .$$ 
{\vskip -.1in}
Thus using the optimal boundary kernel
{\em guarantees the continuity of the estimate}
if at the touch point we apply the optimal interior kernel
of the optimal halfwidth.

At the discontinuity, ($f=f_{disc}$, $\ft = -1$,
$h=h_0(0)$),
the kernel has a simple expression:
$G(-1,\zt) \; = \; P_q(\zt)
\; - \; (2q+3) P_{q+1}(\zt)
\; + \;  (2q+2)  P_{q+2}(\zt) \ .$
The leading order EASE at the boundary is
{\em exactly $4(q+1)^2$ times larger than for the optimal interior kernel}.

One method of constructing kernel shapes is to perform a local 
polynomial regression (LPR) in the neighborhood of $f$. We  model
$\th(f)$ as a $p$th order polynomial: 
$\th(f_i):= \sum_{j=0}^{p-1} a_j(f_i-f)^j$ in the region
$f_i \in [f_{disc}, f_{disc} +2h]$. The $p$ free parameters $\{a_j\}$ 
are chosen by minimizing
{\vskip -.28in}
$$
 F(a_0,a_1,\ldots,a_{p-1}) \; = \;
 \sum_i w_i(f) 
 \left(\sum_{j=0}^{p-1} a_j(f_i-f)^j - \th_i \right)^2 \ .
$$
{\vskip -.08in}
We take $q!a_q$ as the estimate of $\th^{(q)}(f)$.
The weighting functions, $w_i(f)$, are arbitrary and we choose them
to minimize the EASE. The equivalence of LPR and kernel smoothing is given
by

\noindent {\bf Theorem.}
{\em
A kernel of type $(q,p)$ is the equivalent kernel of
local polynomial regression of order $p-1$ with non-negative weights
if and only if the kernel has no more than $p-1$ sign changes.
}

It is known (M\"uller (1987), Fan(1993))
that the optimal interior kernel
of type $(q,p)$,
$p-q\equiv 0\,{\rm mod}\, 2$,
in the continuum limit,
is produced by the scaling weight function $W(y)=1-y^2$.
This choice is not unique!

\noindent {\bf Theorem}
{\em
Let $p-q$ be even.
If data points, $f_i$, in the interval of support, $[f-h,f+h]$,
are symmetric around the estimation point, $t$,
and their weights are chosen as $w_i=W\left(\frac{f_i-f}{h}\right)$,
then each of the functions
$W_1(z)=1-z$, $W_2(z)=1+z$, $W_3(z)=1-z^2$
produces the same estimator
}[6].

Because of the optimality in the interior,
the Bartlett-Priestley weighting, $W(z)=1-z^2$,
is used often in the boundary region as well.
This does not minimize the EASE [6]:

\noindent {\bf Theorem.}
{\em
The asymptotically optimal kernel
is equivalent to a} $\underline{linear}$, {\em 
nonnegative weighting  function. At the boundary,
the equivalent weighting equals $2h-(z-f_{disc})$.
For the intermediate estimation points, $f_{disc}<f<f_{tp}$,
the slope of the weighting line varies as $t$ changes.
For $q=0$, the equivalent weighting is  }
$(1-\ft^2)h+\left( \ft+\sqrt{1-3\ft^2+3\ft^4}\right)h\zt$.
{\vskip -.08in}
\sect{6. 
Summary}
{\vskip -.08in}
We have analyzed the expected asymptotic square error of the
smoothed
log multitapered periodogram and shown that multitapering reduces the
error by a factor of $[{\pi^2 \over 4}]^{.8}$ for the sinusoidal tapers.
The optimal rate of presmoothing prior to taking logarithms is $K \sim
N^{8/15}$, but the expected loss depends only weakly on $K$ when
$1 \ll K \ll Nh$.

We have proposed a data-adaptive multiple stage variable halfwidth kernel
smoother. It has a relative convergence of $N^{-2/9}$, which can be improved
to $N^{-1/4}$ if desired. The multiple stage estimate has the following steps.
0) Estimate $\hat{\theta}_{MT}(f)
\equiv \ln [\hat{S}_{MT}(f)]-B_K /K$ as described in Sec.~2.  
1) Estimate the optimal kernel halfwidth for a kernel of (0,4) 
for the log-single tapered periodogram.
2) Estimate $\theta^{\prime\prime}(f)$ using a kernel smoother of order
(2,4). 3) Estimate $\theta (f)$ using a kernel smoother of order (0,2) with
the halfwidth $h_0 (f)$. The halfwidth is the evaluation of the
asymptotically optimal halfwidth: $h(f) \sim c| \partial^2_f \theta |^{-2/5}
N^{-1/5}$.
\Acknowledgments{Work funded by U.S. Dept.\ of 
Energy Grant DE-FG02-86ER53223}
{\vskip-.16in}
\References


\ref M\"uller, H.,  and U.~Stadtm\"uller,
Variable bandwidth kernel estimators of regression curves,
{\em Annals of Statistics} {\bf 15} (1987) 182--201.

\ref
Riedel, K.~S.,
Kernel estimation of the instantaneous frequency,
{\em I.E.E.E Trans. on Signal Processing} {\bf 42} (1994), 2644-9.


\ref Riedel, K.~S., and A. Sidorenko,
Data Adaptive Kernel Smoothers- How much smoothing?
{\em Computers in Physics} 
{\bf 8} (1994) 402--409.

\ref Riedel, K.~S., and A. Sidorenko,
Minimum bias multiple taper spectral estimation,
{\em I.E.E.E. Trans. on Signal Processing}
{\bf 43}  (1995) 188-195.

\ref Riedel, K.~S., and A. Sidorenko,
Smoothed  multiple taper spectral estimation with data adaptive implemetation,
Submitted. 

\ref  Sidorenko, A. and K.~S.~Riedel, 
Optimal boundary kernels and weightings,
Submitted.

\medskip
\leftline{\it A. Sidorenko}
\leftline{New York University, Courant Institute} 
\leftline{251 Mercer St., New York, NY 10012}
\leftline{\fab sidorenk@cims.nyu.edu}
\medskip
\leftline{\it K.S. Riedel}
\leftline{New York University, Courant Institute} 
\leftline{251 Mercer St., New York, NY 10012}
\leftline{\fab riedel@cims.nyu.edu}


\end{document}